\documentclass[12pt]{iopart}

\usepackage{cite}   
\usepackage{graphicx}
\DeclareGraphicsExtensions{.pdf,.eps}
\usepackage{epsfig}
\usepackage{iopams}
\usepackage{amstext}
\usepackage{hyperref}

\newcommand{\Vb}{\text{V}_{\text{bias}}}

\begin{document}

\title{Light Emission from Ag(111) driven by Inelastic Tunneling in the Field Emission Regime}

\author{Jes\'us Mart\'inez-Blanco$^{1}$, Stefan F\"olsch$^{1}$}
\address{$^{1}$ Paul--Drude--Institut f\"ur Festk\"orperelektronik, Hausvogteiplatz 5-7, 10117, Berlin, Germany}
\ead{jmb.jesus@gmail.com}


\begin{abstract}
We study the light emission from a Ag(111) surface when the bias voltage on a scanning tunneling microscope (STM) junction is ramped into the field emission regime. Above the vacuum level, scanning tunneling spectroscopy (STS) shows a series of well defined resonances associated with the image states of the surface, which are Stark shifted due to the electric field provided by the STM tip. We present photon--energy resolved measurements that unambiguously show that the mechanism for light emission is the radiative decay of surface localized plasmons excited by the electrons that tunnel inelastically into the Stark shifted image states. Our work  illustrates the effect of the tip radius both in the STS spectrum and the light emission maps by repeating the experiment with different tips. 
\end{abstract}

%
\vspace{2pc}

%
\submitto{\JPCM}
%
\maketitle
%
%

\markboth{Light Emission from Ag(111) driven by Inelastic Tunneling...}{Light Emission from Ag(111) driven by Inelastic Tunneling...}

\section{Introduction}

When the bias voltage in an STM junction is ramped above the vacuum level of the surface under study (field emission regime), light is emitted following an oscillatory pattern in intensity as a function of the bias, as reported already in the earliest studies of photon--emission STM \cite{Coombs1988a}. This light emission has been associated with the radiative decay of surface plasmons excited in the tip-surface cavity. Two main mechanisms were proposed for the creation of such plasmons in the field emission regime: hot electron injection and inelastic electron tunneling into field emission states \cite{Berndt1991,Berndt1993b}. These electronic states are confined in the tip--surface gap and correspond to the \textit{image states} of the surface\cite{Echenique1978,Echenique1989,Echenique2002} which are Stark shifted above the vacuum level due to the electric field between the two electrodes of the STM junction\cite{Binnig1985}. By means of STS they can be accessed when the bias voltage between tip and sample is ramped in the field emission regime, something that typically is done with the feedback loop switched on, that is, changing the tip height so that the detected current remains constant. Due to this continuous change of the potential barrier along the spectrum, and the fact that the electrons injected into these states relax interacting with bulk states, they are detected as field emission resonances (FER in the following). The existence of these states has been exploited in the past for many diverse purposes, like obtaining chemical contrast \cite{Jung1995}, measuring work function fluctuations \cite{Ploigt2007,Ruffieux2009}, achieving atomically resolved STM images on insulators \cite{Bobrov2001} or even as qubit holders for quantum computation \cite{Platzman1999}. By studying the light emitted from the junction at the field emission regime, we aim to gain insight on the underlying mechanisms producing it and its relationship with the field emission electronic states.

Our results provide direct experimental evidence of plasmon generation on a noble metal surface for bias voltages in the field emission regime, and they show unambiguously that they are created via inelastic tunneling electrons. As discussed in Ref. \cite{Persson1992a}, the hot electron mechanism has a very low efficiency for surface plasmon excitation and therefore the principal excitation process occurs via inelastic tunneling, as demonstrated here. Previous studies addressing the light emission in the field emission regime rely on the acquisition of a particular photon energy with a photomultiplier (isochromatic mode). We have measured the evolution of light intensity as a function of the applied bias voltage acquiring the full spectrum resolved in photon energy. By using the multi--channel instead of the single--channel detection, we get a more complete picture, that allows us to directly compare the situation in the tunneling regime and in the field emission regime. Since both the differential conductance and the light emission spectra depend on the tip geometry, we present results for different tips whose radii are estimated by comparing the experimental FER energies with the calculated confinement energies of an electron in a simple 1D potential defined between a flat surface and a spherical tip. 

\section{Experimental}
The experiments were carried out in an ultra--high vacuum chamber equipped with a low--temperature STM (\textit{Createc GmbH}) operated at 5 K. A clean and atomically flat Ag(111) surface was prepared by repeated cycles of Ne$^{+}$ sputtering and surface annealing. We use a standard etched tungsten STM tip prepared in vacuum by front sputtering and annealing. Once the tip is inserted in the STM scanner, we condition it in situ either by soft indentation on the sample (a few nanometers) or by deep indentation (a few hundreds of nanometers) while applying a bias voltage of -80 V to the sample and subsequently withdraw the tip until the junction is broken. 

The photons generated at the STM junction are collected by a lens installed at a distance of 4.5 cm, and focused by another lens outside the chamber into an optical fibre. The light is then guided to the entrance slit of a spectrometer (\textit{SpectraPro 2500i, Princeton Instruments}) equipped with a 150 grooves/mm grating element and a liquid nitrogen cooled CCD camera. The detection efficiency of the acquisition setup, including all elements from the junction to the CCD camera, is plotted in figure \ref{Plasmons}(b) as a function of the photon energy. This efficiency curve is provided for reference, and has not been used to correct the luminescence signal shown in the figures. The detection axis resides at an angle of 14$^{\circ}$ with respect to the surface plane. The aperture of light collection amounts to 0.2 steradian. During the experiments, the light was measured as the bias voltage was ramped in steps of 100 mV, with acquisition times of 2 minutes per step. The spectra were acquired with the feedback loop switched on, so that the tip height is changing to keep the current constant.

\section{Results and Discussion}
\begin{figure}
\begin{center}
\includegraphics[width=0.8\columnwidth]{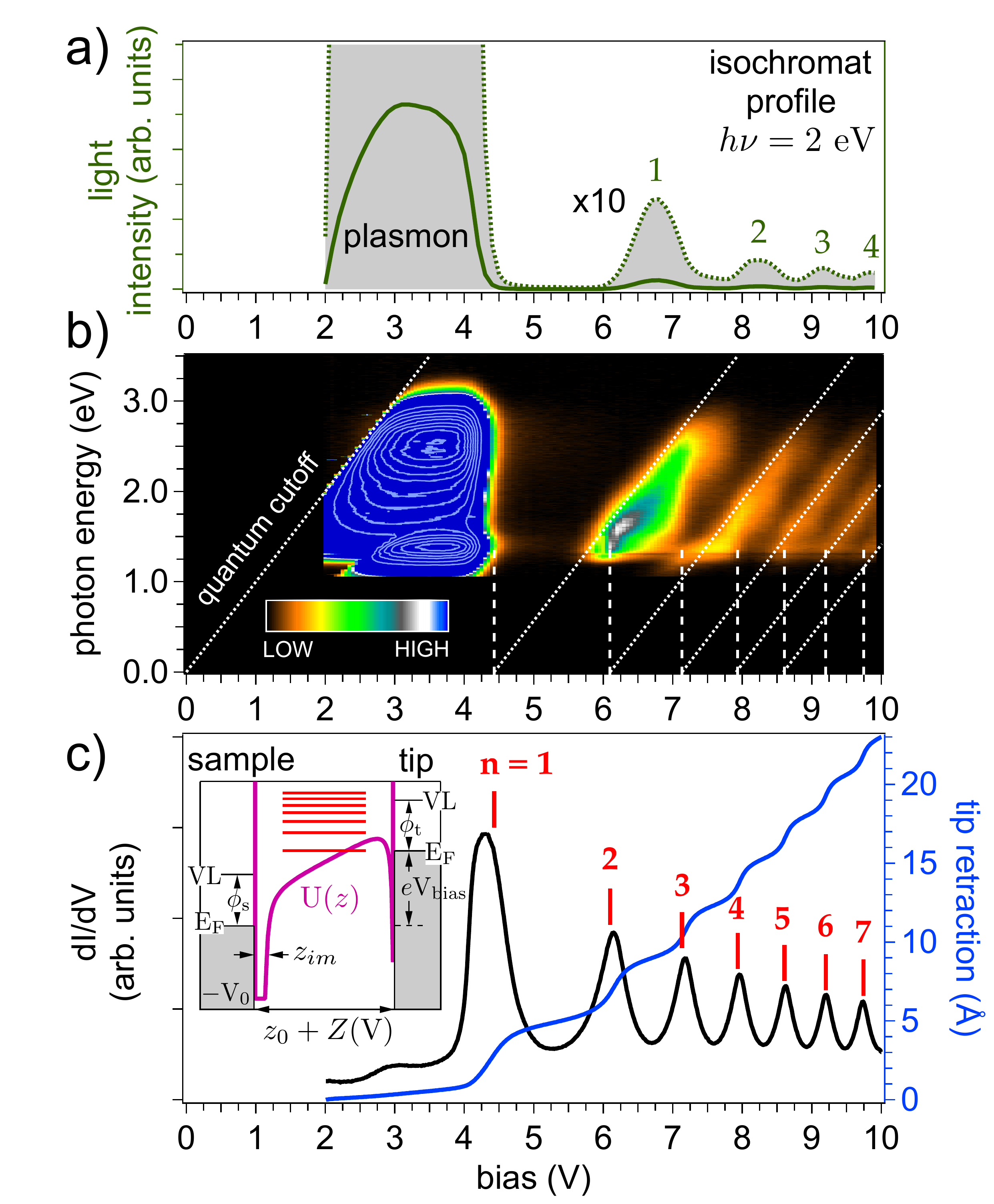}
\end{center}
\caption{(a) Profile of the light intensity map shown in (b) along the isochromat line corresponding to h$\nu = 2$ eV for bias voltages between 2 and 10 V. The dotted line represents the same profile scaled by a factor of 10 for better visualization. (b) Light intensity map as a function of bias and photon energy, measured at constant current (200 nA). The color scale has been replaced by a contour plot in the region 2--4 eV due to the high plasmon related intensity. Dotted white lines mark the quantum cutoff (bias=photon energy) and the corresponding shifted cutoffs due to inelastic scattering into the different FER. The FER energies are marked with vertical dashed lines. (c) In black, the differential conductance as a function of the bias voltage at constant current (200 nA). In blue, the tip retraction with respect to the initial tip--sample distance, acquired simultaneously. The vertical red lines indicate the simulated energies of the first seven FERs according to the model depicted in the inset (see text for details).}	
\label{FER_simulation}
\end{figure}

The black curve in Fig.\ref{FER_simulation}(c) corresponds to the differential conductance signal measured on Ag(111) at a constant current of 200 nA as the bias voltage was ramped from 2 to 10 V. The tip retraction was simultaneously acquired and is represented in blue in this panel. From 4 to 10 V bias, 7 prominent peaks are clearly visible, and identified as the first FERs. With a simple model potential, it is possible to reproduce the binding energies at which the different FERs appear in the STS experiment \cite{Ploigt2007}. The proposed one--dimensional potential U($z$) is depicted in the inset of Fig.\ref{FER_simulation}(c) and it is defined along the $z$ axis from a silver surface to a spherical tip of radius $R$. The calculated energies are shown as vertical red lines and are found by using the Numerov method \cite{Numerov1927} to integrate numerically\cite{Martinez-Blanco2015} the time--independent Schr\"odinger equation for an electron confined in the potential obtained by adding three different contributions:

\begin{enumerate}
\item{The surface image potential,
\begin{equation} 
   \frac{-1}{4\pi\epsilon_0}\frac{e^2}{4}\left(\frac{1}{z-z_{im}}\right),
   \label{SIP}
\end{equation} except for the integration points near the image plane $z=z_{im}$, in which the potential is constant and equal to the crystal inner potential $-\text{V}_0$ to avoid singularities. In our case, we take the minimum potential in the bulk of silver (-10 eV with respect to the vacuum level VL of the surface) as a suitable value for the inner potential \cite{Ploigt2007}. 
}
\item{The tip image potential,
\begin{equation} 
   \frac{-1}{4\pi\epsilon_0}\frac{e^2}{4}\left(\frac{1}{z_0+Z(\text{V})-z}\right),
   \label{TIP}
\end{equation}
}
\item The electric potential between sample and tip. Assuming a spherical tip, the value of this potential along the cylindrical symmetry axis (coordinate $z$) can be calculated according to Ref. \cite{DallAgnol2009} and will depend on the effective radius R of the tip, the bias voltage and the tip--sample separation $z_0+Z(\text{V})$.
\end{enumerate}

\begin{figure}
\begin{center}
\includegraphics[width=0.75\columnwidth]{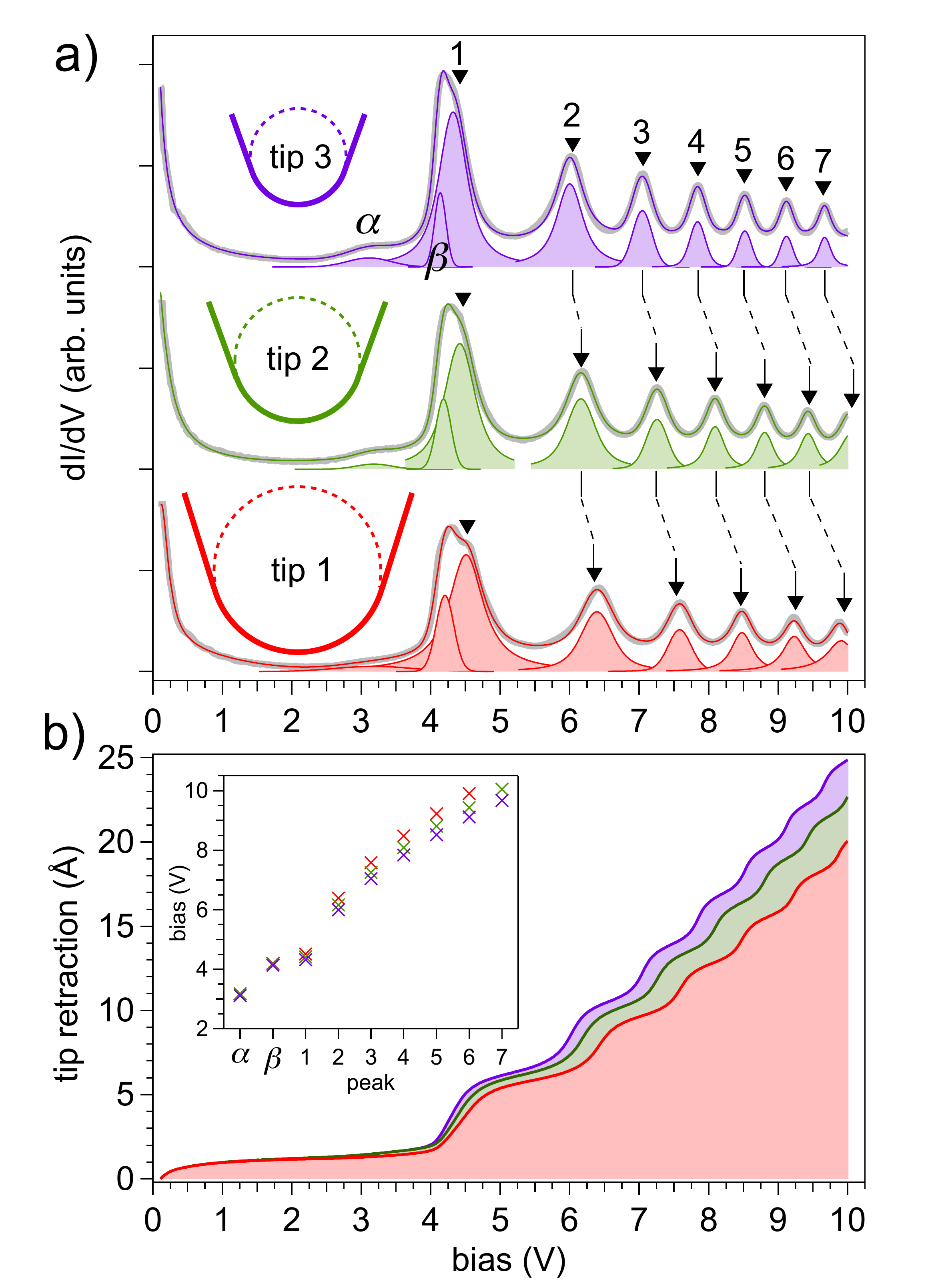}
\end{center}
\caption{(a) In light grey, the experimental differential conductance measured at constant current (50 nA) with three different tips on Ag(111). The corresponding red, green and violet profiles are the result of fitting the data to peaks with Voigt profiles on top of a polynomial background. The triangles mark the bias voltages at which the simulated FER peaks are to be found. The simulation uses a tip radius of 42 {\AA}, 54 {\AA} and 91 {\AA} for tip 1, 2 and 3 respectively. (b) The simultaneously measured tip retractions for tips 1, 2 and 3. Larger tip radius corresponds to shorter retractions, in accordance to Ref. \cite{Pitarke1990}. The inset shows the bias position of the different peaks of the spectra for the three tips used in the study.}
\label{TIP_dependence}
\end{figure}

The potential is considered as infinite at both sides of the vacuum gap, for which we neglect the bulk potential in this approximation. In expression (\ref{SIP}), $e$ is the electron charge and $\epsilon_0$ is the vacuum permitivity. In expression (\ref{TIP}), $Z(\text{V})$ is the experimental tip displacement (blue curve in Fig.\ref{FER_simulation}(c)) with respect to the initial tip--sample distance $z_0$. The value of $z_0$ and $z_{im}$ is taken from the fit described below for Fig.\ref{TIP_dependence}, and only the effective tip radius $R$ has been adjusted to reproduce the position of the experimental peaks, with a final value of 40 {\AA}. Since the potential barrier is changing continuously as the bias voltage $\Vb$ is ramped up (due to the tip retraction), we compute the list of eigenenergies $E_n$ of the system with respect to the vacuum level of the surface VL for each $\Vb$, and then we extract the bias voltage at which we expect to encounter the FER $n$
with the condition $E_n(\Vb)=E_{\text{F,tip}}=E_{\text{F,sample}}+e\Vb=-\phi_\text{s}+e\Vb$, where $E_{\text{F}}$ is the Fermi energy and $\phi_\text{s}$ is the work function of the surface.

The color map of Fig.\ref{FER_simulation}(b) shows the detected light resolved in photon energy emitted by the junction as the bias voltage is ramped from 2 to 10 V and the tunneling current is kept constant and equal to 200 nA. Since this experiment was done at the same surface spot and with the same tip used for the STS experiment, we can directly correlate the position of the FERs and the oscillations observed in the light map. The vertical white dashed lines in Fig.\ref{FER_simulation}(b) mark the FER positions encountered in the STS experiment and the white dotted line on the left mark the quantum cutoff, the line delimiting the region at which photons with energy $h\nu \leq e\Vb$ are possibly emitted. The high intensity detected from 2 V up to $\sim 4.3$ V corresponds to the well known radiative decay of a surface localized plasmon. As the bias voltage approaches $\sim 4.3$ V, the tunnel junction enters in resonance with the first FER and the STM feedback loop has to increase drastically the tip--surface distance to keep the tunneling current constant. Eventually, this distance becomes large enough to weaken the tip--surface electromagnetic coupling and plasmons are no longer generated, for which light emission is quenched\cite{Johansson1990}. The light intensity peak observed at $\sim 3.5$ V bias and with $\sim 1.3$ eV photon energy is an artifact and due to the second order refraction of the spectrometer grating.

As the bias voltage is ramped further up into the field emission regime, light is again detected above a cutoff defined by $h\nu=e\Vb-E_1$, where $E_1$ is the energy of the first FER ($n=1$). The light intensity map clearly shows that the pattern is repeated at higher bias voltages and photons are successively detected above cutoffs defined by the condition $h\nu=e\Vb-E_n$, with $n=1,2,3,4,...$ (white dotted lines). This light is originated again in the radiative decay of the surface localized plasmon, which is now excited due to the energy lost by the electrons that inelastically tunnel from the Fermi level of the tip into the FER states. The mechanism is similar to the one described in Ref.\cite{Hoffmann2001} for the luminescence from a quantum well system on a metallic surface.

It is worth noting that the intensity distribution at these higher voltages is modulated with a reduction every time the junction enters in resonance with a FER (vertical dashed lines). This might be a consequence of two different effects. On the one hand, at these particular bias voltages the tip--surface distance gets larger in a step-like fashion, limiting the amplitude of the generated plasmon\cite{Johansson1990} due to an abrupt change in the cavity separation. On the other hand, as the junction resonates with a new FER state, another tunneling channel is available and thus there are less electrons exciting the plasmon through the original one.

In order to better visualize the intensity contrast of the different processes, the graph presented in Fig.\ref{FER_simulation}(a) shows an isochromat profile taken from panel (b) of this figure at $h\nu=2$ eV. The plasmon related intensity from inelastic tunneling into the first FER (marked with the number 1) is about 20 times less intense than the plasmonic intensity detected in the tunneling regime, and gets successively smaller for the next FERs.

\begin{figure}
\begin{center}
\includegraphics[width=1\columnwidth]{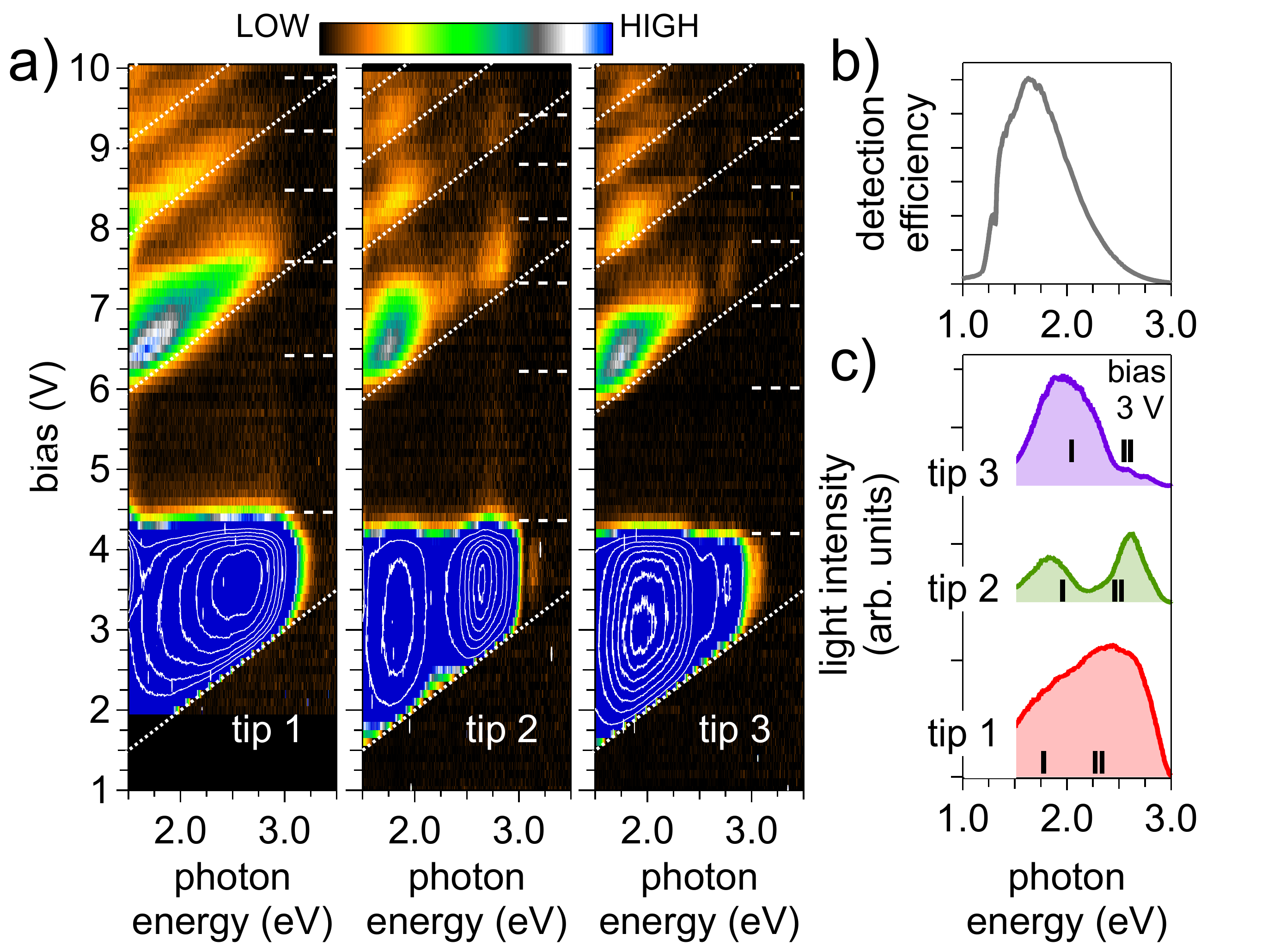}
\end{center}
\caption{(a) Light intensity maps measured on Ag(111) using the three tips described in Fig. \ref{TIP_dependence}. The horizontal dashed lines mark the position of the FER peaks found in the STS experiment. The dotted white lines mark the quantum cutoff (bias=photon energy) and the corresponding shifted cutoffs due to inelastic scattering into the different FER. The contour plots are used to better visualize the energy distribution of the plasmon originated light intensity. (b) Detection efficiency spectrum of our acquisition device showing that our sensitivity is significant for photon energies between $\sim 1.3$ and $\sim 2.5$ eV. (c) Light intensity profiles from the maps of panel (a) for a bias voltage of 3 V. The vertical lines mark the energy of the first three modes of the localized surface plasmons as calculated using the plane-sphere model \cite{Rendell1978,Rendell1981} with the tip radii extracted from the STS simulation described in Fig. \ref{TIP_dependence}.}
\label{Plasmons}
\end{figure}

By repeatedly conditioning the STM tip in the way explained in the experimental section, we get different tip geometries that give rise to significant changes in the STS spectra as shown in Fig.\ref{TIP_dependence} as well as in the corresponding luminescence maps (shown later in Fig.\ref{Plasmons}). Fig.\ref{TIP_dependence}(a) presents the experimental differential conductance at constant current (50 nA) for three different tips. The spectra are displayed assuming an increasing tip radius from top to bottom. This assumption is based on the different overall response of the tip retraction, which is larger for smaller tip radius as described in Ref. \cite{Pitarke1990}. The experimental tip retraction with the bias voltage is shown in Fig.\ref{TIP_dependence}(b) for each of the three tips, and we note that the violet spectrum (corresponding to the tip labelled with the number 3) shows a total tip retraction which is $\sim 5$ {\AA} larger than the spectrum for tip number 1.

This trend assignment is further confirmed by the simulation of the FER positions following the procedure described previously for Fig.\ref{FER_simulation}. The black triangles in Fig.\ref{TIP_dependence}(a) mark the calculated bias voltages at which we expect to find the different FERs, using the same set of model parameters for the three tips and changing only the tip radius $R$. Since the tip is very likely to be silver coated, we take as a reasonable approximation the same work function for the tip and for Ag(111) (4.46 eV according to Ref.\cite{Chelvayohan1982}). Considering a silver inner potential V$_0 = -10$ eV as explained earlier, there are only two free parameters in the model potential besides the radius of the tip. These are the starting tip--surface separation $z_0$ and the image plane $z_{im}$. The best combined fit of the experimental FER positions was found for $z_0 = 14$ {\AA} and $z_{im} = 1.9$ {\AA} and the following tip radii: $R_{\text{tip 1}}=91$ {\AA}, $R_{\text{tip 2}}=54$ {\AA} and $R_{\text{tip 3}}=42$ {\AA}.

In the previous analysis, the experimental peak positions were obtained by fitting the dI/dV spectra to a series of Voigt profiles on top of a polynomial background. As indicated in Fig.\ref{TIP_dependence}(a), in order to account for the anomalous peak shape of the first FER (specially for tip 1 that most evidently appears as a multiple peak resonance), the fit was performed using three peaks, labelled as $\alpha$, $\beta$ and 1 in the figure. Looking at the inset of Fig.\ref{TIP_dependence}(b) we observe that the bias position of the FER peaks (labelled with quantum numbers 1, 2, 3,...) are the only ones that shift for different tips. Peaks $\alpha$ and $\beta$ are located near the first FER but they should have a different origin. Interestingly, STS studies performed on the (111) faces of other noble metals like Cu(111) and Au(111) reveal also multipeak features at the bias voltages at which the first FER is encountered \cite{Dougherty2007}. For Au(111), k-resolved inverse photoemission studies\cite{Woodruff1986} detect a resonance near the first FER which is assigned to the edge of the bulk band projection. Similarly, Refs.\cite{Hulbert1985,Goldmann1985} report for Ag(111) a band edge position of about 4 eV above the Fermi level, which is compatible with the position of peak $\beta$ in our STS spectra. Finally, feature $\alpha$ can be considered a shoulder rather than a peak and has not been observed for STS experiment at lower tunneling currents\cite{Sperl2008}.

The color maps of Fig.\ref{Plasmons}(a) show the light intensity collected for each of the three tips described in Fig.\ref{TIP_dependence} during the corresponding STS experiment at 50 nA. As we can directly see, the size of the tip has a strong impact on the detected light distribution. In particular, the cutoffs found in the field emission regime are distributed in good agreement with the condition $h\nu=e\Vb-E_n$ (white dotted lines), where $E_n$ follow a different series of FER energies for each tip (horizontal dashed lines). As in the experiment shown in Fig.\ref{FER_simulation}, we observe a systematic decrease of intensity at each of these dashed lines.

Comparing the three color maps of Fig.\ref{Plasmons}(a) we observe as well a large effect of the tip on the photon energy distribution of the plasmon originated signal, displayed here with a white contour plot due to its high intensity with respect to the field emission intensity. The differences are more clearly seen in the line profiles of Fig.\ref{Plasmons}(c) extracted from these maps at 3 V bias voltage. An observable influence of the tip on the light emission spectra has been reported before \cite{Rossel2009}, and actually it can be used to conveniently tune the nanocavity plasmon resonance mode \cite{Zhang2013}. It is well established that the spectral distribution of the emitted light can change largely with the tip geometry as described in Ref. \cite{Aizpurua2000} for a hyperboloid tip model. Here, we will use the sphere tip model of Refs. \cite{Rendell1978,Rendell1981} to calculate the photon energies at which we expect to have localized surface plasmon modes in the luminescence experiment\cite{Johansson1990,Johansson1991,Downes1998a}, since we already have an estimation for the main parameter of this model, namely the radius $R$ of the tip. According to this model and assuming a free electron gas behavior for the valence electrons of the metal substrate, the resonant frequency ($\omega_{\text{LSP}}$) of the surface localized plasmon mode $n$ can be written in terms of the bulk plasmon frequency of the metal substrate ($\omega_{\text{P}}$) as\cite{Rossel2010}:

\begin{equation} 
	\omega_{\text{LSP}}=\omega_{\text{P}}\sqrt{\frac{\tanh{\left(n+\frac{1}{2}\right)} \beta_0}{\epsilon_0+\tanh{\left(n+\frac{1}{2}\right)} \beta_0}},
   \label{plasmonfreqs}
\end{equation}

where $\beta_0=\cosh^{-1}(1+d/R)$ and $d$ is the tip--surface distance. The vertical black lines of Fig.\ref{Plasmons}(c) mark the position of the first three plasmon modes calculated according to (\ref{plasmonfreqs}). A value of 15 {\AA} is used for $d$ (obtained from the expected tip--surface separation at 50 nA and 3 V as shown in Fig.\ref{TIP_dependence}) and for $R$ we use the corresponding tip radius obtained in the fitting described above. Finally, the bulk plasmon energy used for the silver substrate is $\hbar\omega_{\text{P}}=3.9$ eV. This value is considerably smaller than the commonly accepted value of $\sim 9$ eV (see Ref. \cite{Zeman1987}), and yet there are reasons to consider it realistic as described in Ref.\cite{Boyle2009}. 

As can be seen in Fig.\ref{Plasmons}(c), the calculated plasmon energies shift to lower values as the tip radius increases\cite{Suzuki1999}. This trend is compatible with the observed spectra, but a direct identification of the experimental plasmonic modes is more challenging. Specifically, the detected spectral features are not always in direct correspondence with the emitted spectrum. Apart from the finite response of the detector (shown in Fig.\ref{Plasmons}(b)), which obviously will make undetectable a plasmon mode if its energy is too high or too low, we also have the effect of the local topography of the tip apex. For a given tip radius, the details of the apex at the local level, which are totally unknown, may have an influence on the appearance of the resulting plasmonic spectrum\cite{Meguro2002}. Nevertheless, regardless of the specific spectral features obtained for a particular tip in the tunneling regime, we observe a consistent repetition of those features in the field emission regime above the different cutoffs. 

\section{Summary}
We present comprehensive luminescence maps resolved in photon energy showing the light intensity distribution emitted by Ag(111) at an STM junction. The light spectra are correlated with constant current scanning tunneling spectra for bias voltages from the tunneling regime to the field emission regime. The light emission at the field emission regime is explained by the decay of localized surface plasmons which are excited by electrons that inelastically tunnel from the Fermi level of the tip into the Stark shifted image states of the surface. The binding energy of these states is different for different tip sizes and the corresponding light intensity distribution changes accordingly. The radius of the different tips used here are estimated by fitting the position of the experimental field emission resonances to the binding energies expected for an electron confined between a flat surface and a spherical tip. Our results show full consistency between the expected trend in the change of electron and photon spectroscopies for different tips, and confirms the inelastic tunneling mechanism for the emission of light in the field emission regime.

\ack This work was supported by the German Research Foundation (SFB 658, TP A2).

\section*{References}

\providecommand{\newblock}{}

\end{document}